\documentclass[10pt]{article}
\usepackage[dvips]{graphicx}
\usepackage{float}
\usepackage{setspace}

\begin{document}
\doublespacing

\begin{titlepage}

\begin{center}
  {\Large\bf 
  
  Point Process Analysis of Vortices in a Periodic Box\\}

\

\

  {\large Makoto Umeki}
  
  {\large Department of Physics, Graduate School of Science}

  {\large University of Tokyo, 7-3-1 Hongo, Bunkyo-ku, Tokyo 113-0033}

\end{center}

\ \ 

\begin{abstract}

\normalsize

The motion of assemblies of point vortices in a periodic parallelogram 
can be described by the complex position $z_j(t)$ whose 
time derivative is given by the sum of the complex velocities 
induced by other vortices and the solid rotation centered at $z_j$. 
A numerical simulation up to 100 vortices in a square periodic box 
is performed with various initial conditions, including single and double 
rows, uniform spacing, checkered pattern, and complete spatial randomness. 
Point process theory in spatial ecology is applied in order to 
quantify clustering of the distribution of vortices.
In many cases, clustering of the distribution persists 
after a long time if the initial condition is clustered. 
In the case of positive and negative vortices with the same absolute 
value of strength, the $L$ function becomes positive for 
both types of vortices. 
Scattering or recoupling of pairs of vortices by a third vortex 
is remarkable.

\end{abstract}

\end{titlepage}

\section{Introduction}

A motion of point vortices (PVs) is one of the simple and fundamental 
issues in fluid dynamics \cite{Saffman}. 
There have been a number of numerical studies 
of PVs, whose motion is bounded by a circular wall. 
Recently, stable triangular vortex lattices have been 
observed in rotating Bose-Einstein condensates of 
Na atoms \cite{Abo} in which 
the circulation of vortices is quantized with the same strength 
and the lattice contains over 100 vortices. 
Many experimental reports of superfluid vortex lattices 
increased in these several years show triangular patterns \cite{Abo,Kita,Tsubota}, 
which can be compared with the stability analysis of vortex lattices 
by Tkachenko \cite{Tkachenko1,Tkachenko2} in 1966. 
Conformal theory of irrotational flows and complex functions 
are invoked and the role of the Weierstrass zeta function is crucial. 

This paper supplements the Letter \cite{Umeki} by showing the 
spatial distribution of PVs for various type of initial conditions. 
Tkachenko's work 40 years ago and recent strong interests by physicists 
on vortex lattices in Bose-Einstein condensates stimulate 
the author to consider the numerical study of motions of PVs with 
periodic boundary conditions. 

In order to quantify clustering of the point distribution, 
we apply point process theory used in spatial ecology. 
The computed $K$ and $L$ functions judge the distribution 
to be clustered or not. 
This method can be applied to other systems which include 
information of the point distribution in two or more dimensions. 

\section{Velocity field by periodic point vortices}

The velocity field by a single periodic PV of strength $\kappa=2\pi$ 
is given by \cite{Tkachenko1,Aref2,Umeki}
\begin{equation}
w(z)= u+iv= i \overline{\zeta(z)} - i \Omega z, 
\label{eq1}
\end{equation}
where $z=x+iy$ and the overbar denotes the complex conjugate. 
The corresponding streamfunction is denoted by
\begin{equation}
\psi=-{\rm Re} \ln \sigma(z) +\Omega |z|^2 /2.
\label{eq12}
\end{equation}
The Weierstrass zeta and sigma functions ($\zeta$ and $\sigma$) 
contain two complex parameters 
$\omega_1$ and $\omega_2$, which are half periods on the complex plane. 
The second term including 
$\Omega = \pi/[ 4 {\rm Im} (\bar{\omega}_1 \omega_2) ]$ 
expresses the solid rotation in order to cancel the 
circulation on sides of the parallelogram induced by the PV. 

The equations of motions for PVs of strength $2\pi \mu_i$, $\mu_i=\pm 1$, 
can be described by 
\begin{equation}
\dot{z}_i= \sum_{j\ne i} \mu_j w(z_i-z_j).   
\label{eq3}
\end{equation}
For a single type of PVs with the total number $N$, the index $i$ 
is $i=1, \cdots, N$. For positive and negative PVs with the 
same strength, the index is 
$i=1, \cdots, N_{+}, N_{+}+1, \cdots, N(=N_{+}+N_{-})$. 
There are three known conserved quantities; the Hamiltonian and 
two components of the linear impulse \cite{Umeki}. 

\section{K and L functions}

The $K$ function used in spatial ecology \cite{Cressie} is defined by 
\begin{equation}
K(r) = (\lambda N)^{-1}
\sum_{i=1}^N\sum_{j=1,j\ne i}^N
\theta(r - | \vec{x}_i-\vec{x}_j | ), 
\label{eq4}
\end{equation}
where $N$ is the total number of points, $\lambda=N/S$ is the number density, 
$S$ is the area, $\vec{x}_j$ is the position of the $j$th points, 
and $\theta(r)$ is the step function. 
A function for the edge correction 
on the right hand side of (\ref{eq4}), 
which usually appears in data analysis of spatial ecology, 
is unnecessary in the present periodic case. 
The $L$ function is given by 
\begin{equation}
L(r)= \sqrt{K(r)/\pi} -r.
\label{eq5}
\end{equation}
If the distribution of points is complete spatial randomness (CSR), then 
$K=\pi r^2$ and $L=0$. If the points are clustered, $L>0$. 
The uniformly spaced (US) distribution leads to $L<0$. 
It is generally difficult to distinguish between CSR 
and US distributions only if we see raw data of the point 
distribution. However, the $L$ function gives judgment on 
the degree of clustering clearly. 

We note that the value of $L(r)$ depends on $r$. 
For examples, the checkered pattern gives a positive value of 
$L(r)$ for $2 r \simeq l_s$, where $l_s$ is 
the size of the smallest square, but $L(r)<0$ for $4 r \simeq l_s$. 

\section{Numerical simulation of a single type of PVs}

The system (\ref{eq3}) with $N$ up to 100 is simulated numerically by using 
{\it Mathematica} 5.2. 
The textbook by Trott (2006) \cite{Trott} gives 
a number of numerical codes for {\it Mathematica} programming, 
including samples of point vortices with other boundary conditions.

We first consider the following four initial conditions: 
(I) an infinite row that is a discrete model of the vortex sheet, 
(II) PVs located randomly in checkered patterns,  
(III) PVs located randomly in the 10 $\times$ 10 subsquares,
and (IV) CSR in the unit square. 
The typical initial and final distributions of PVs are shown in 
Figures 1-4. The corresponding $L$ functions can be found in \cite{Umeki}. 

\begin{figure}[H]
  \begin{center}
  \vspace{-25mm}
    \begin{tabular}{cc}
      \resizebox{60mm}{!}{\includegraphics{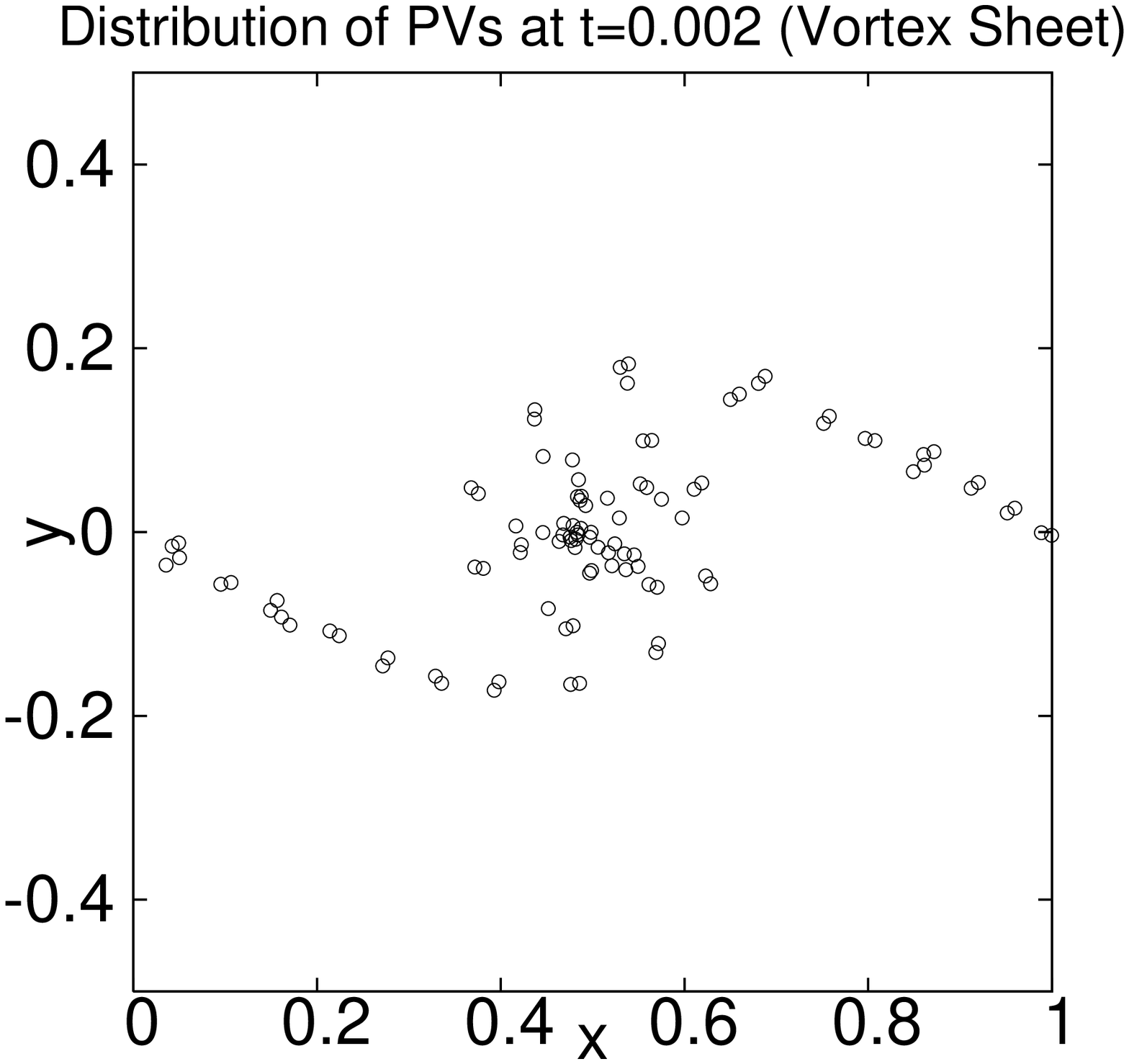}} &
      \resizebox{60mm}{!}{\includegraphics{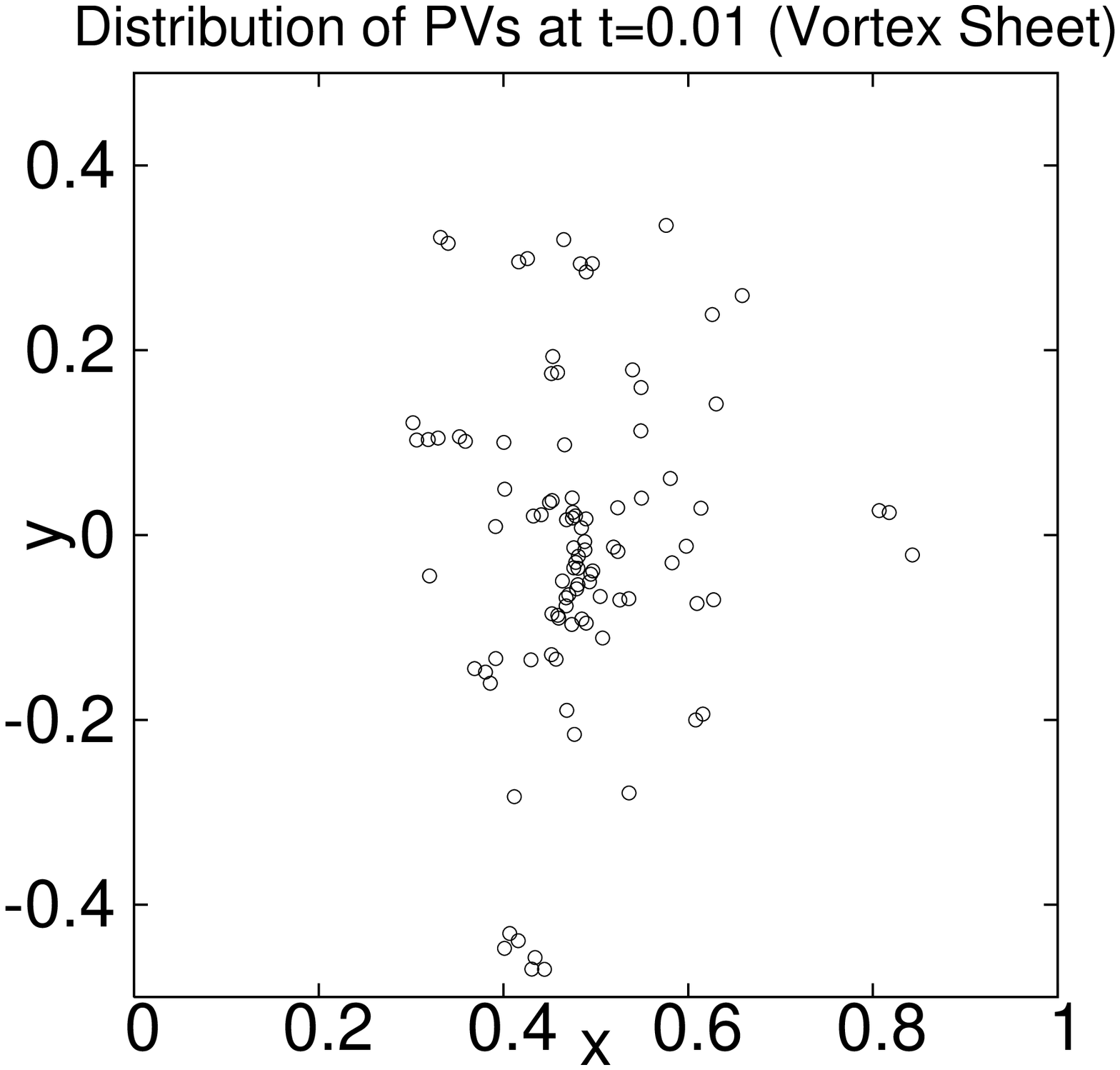}} \\
    \end{tabular}
    \caption{(I) The positions of PVs at $t=0.002$ and $t=0.01$ denoting the 
    discrete approximation of the vortex sheet.}
    \label{fig1}
  \end{center}
  \vspace{-25mm}

  \begin{center}
    \begin{tabular}{cc}
      \resizebox{60mm}{!}{\includegraphics{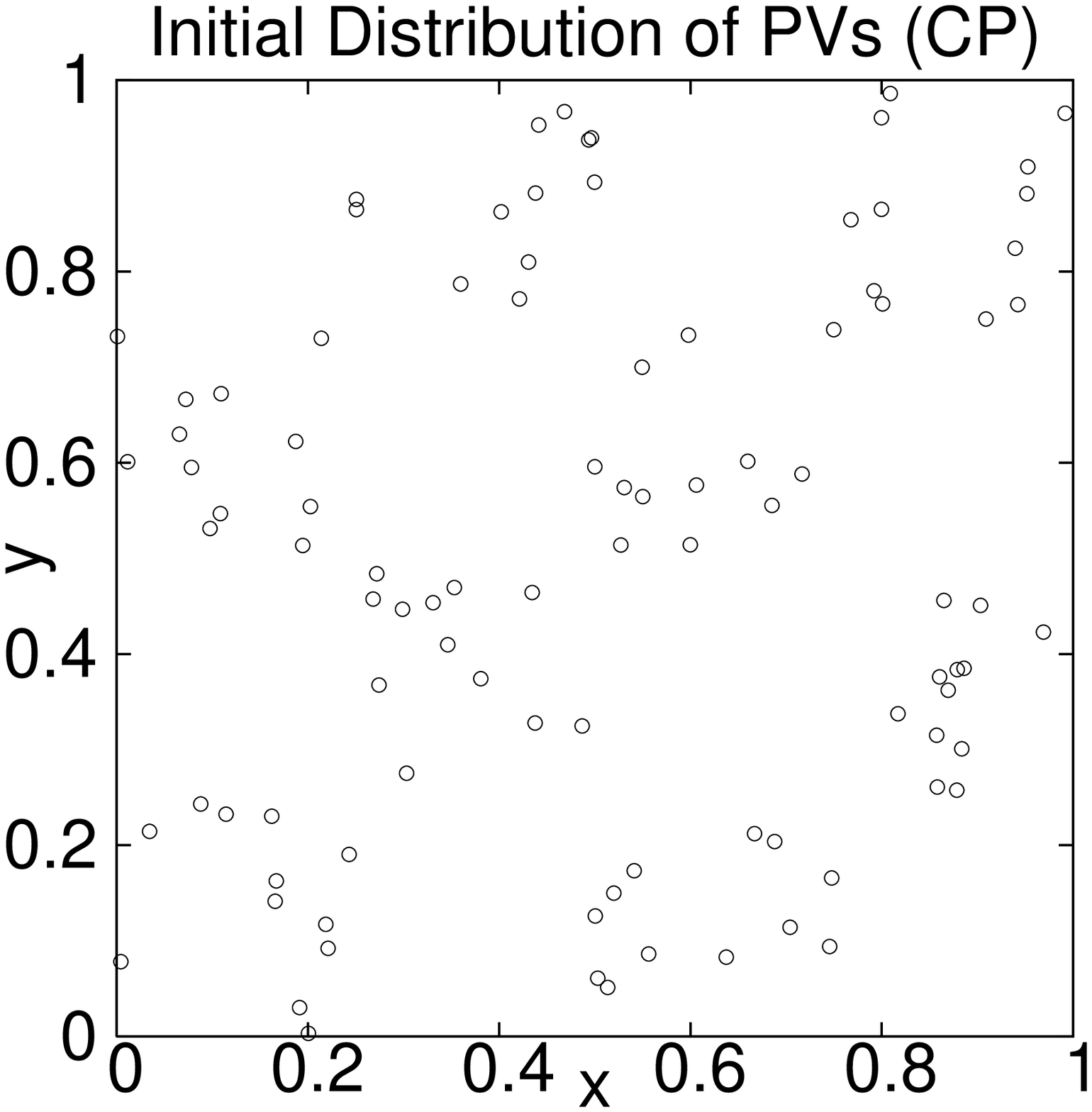}} &
      \resizebox{60mm}{!}{\includegraphics{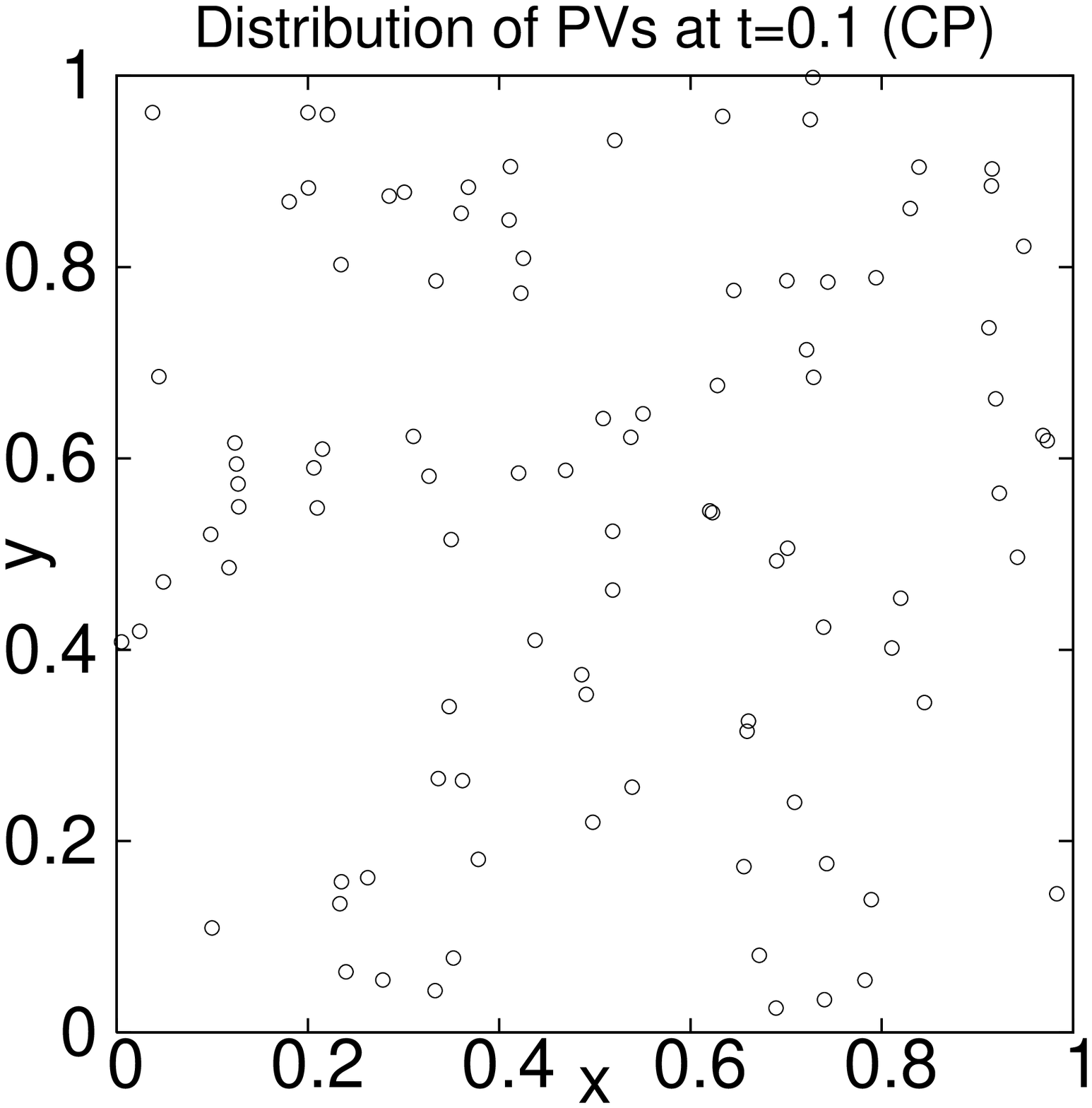}} \\
    \end{tabular}
    \caption{(II) The positions of PVs at $t=0$ and $t=0.1$, 
    denoting the checkered pattern.}
    \label{fig2}
    \end{center}
\vspace{-25mm}
  \begin{center}
    \begin{tabular}{cc}
      \resizebox{60mm}{!}{\includegraphics{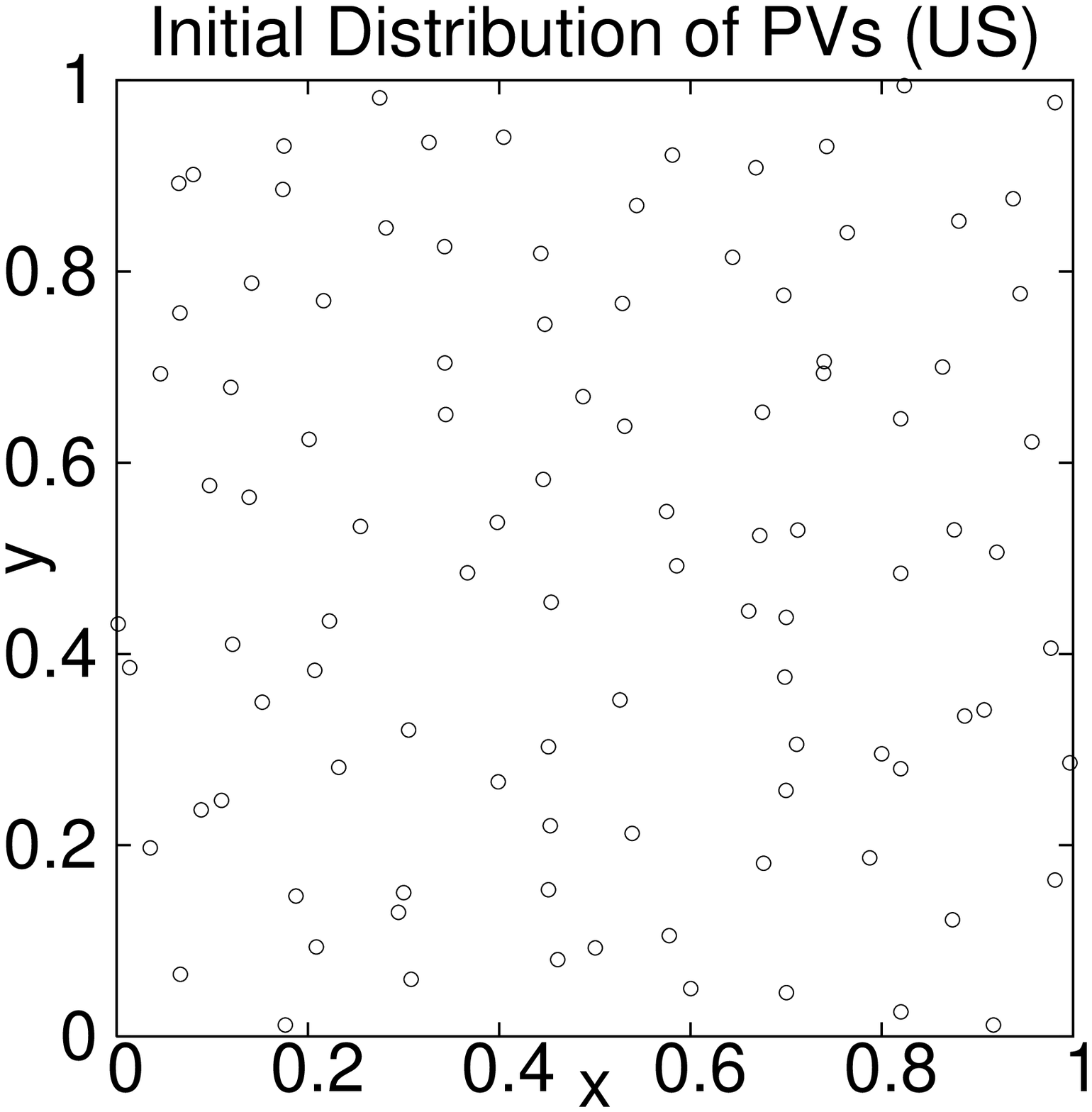}} &
      \resizebox{60mm}{!}{\includegraphics{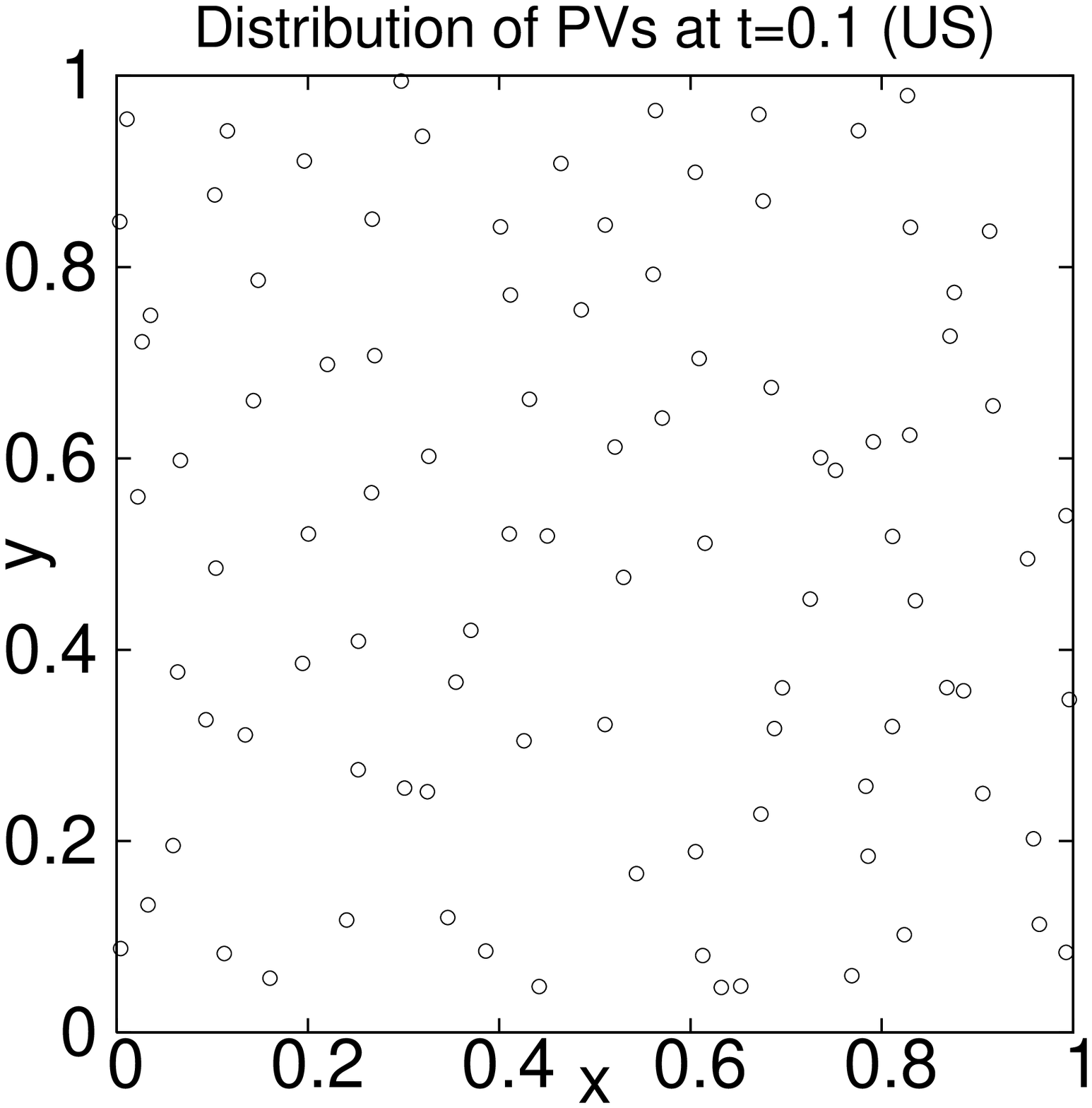}} \\
    \end{tabular}
    \caption{(III) The positions of PVs at $t=0$ and $t=0.1$, 
    denoting the uniformly spaced pattern.}
    \label{fig3}
      \end{center}
\end{figure}
\begin{figure}[H]
  \begin{center}
  \vspace{-25mm}
    \begin{tabular}{cc}
      \resizebox{60mm}{!}{\includegraphics{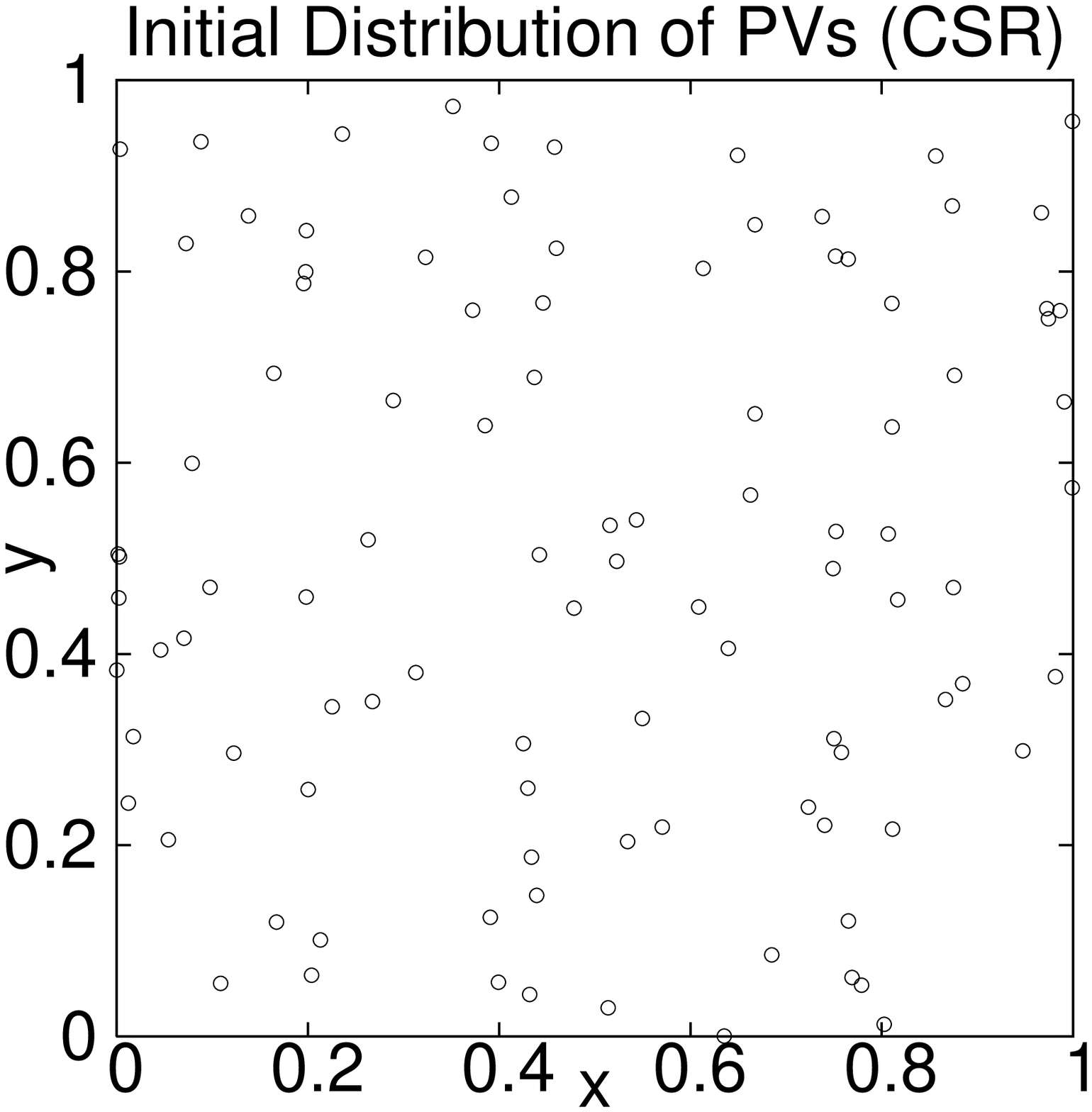}} &
      \resizebox{60mm}{!}{\includegraphics{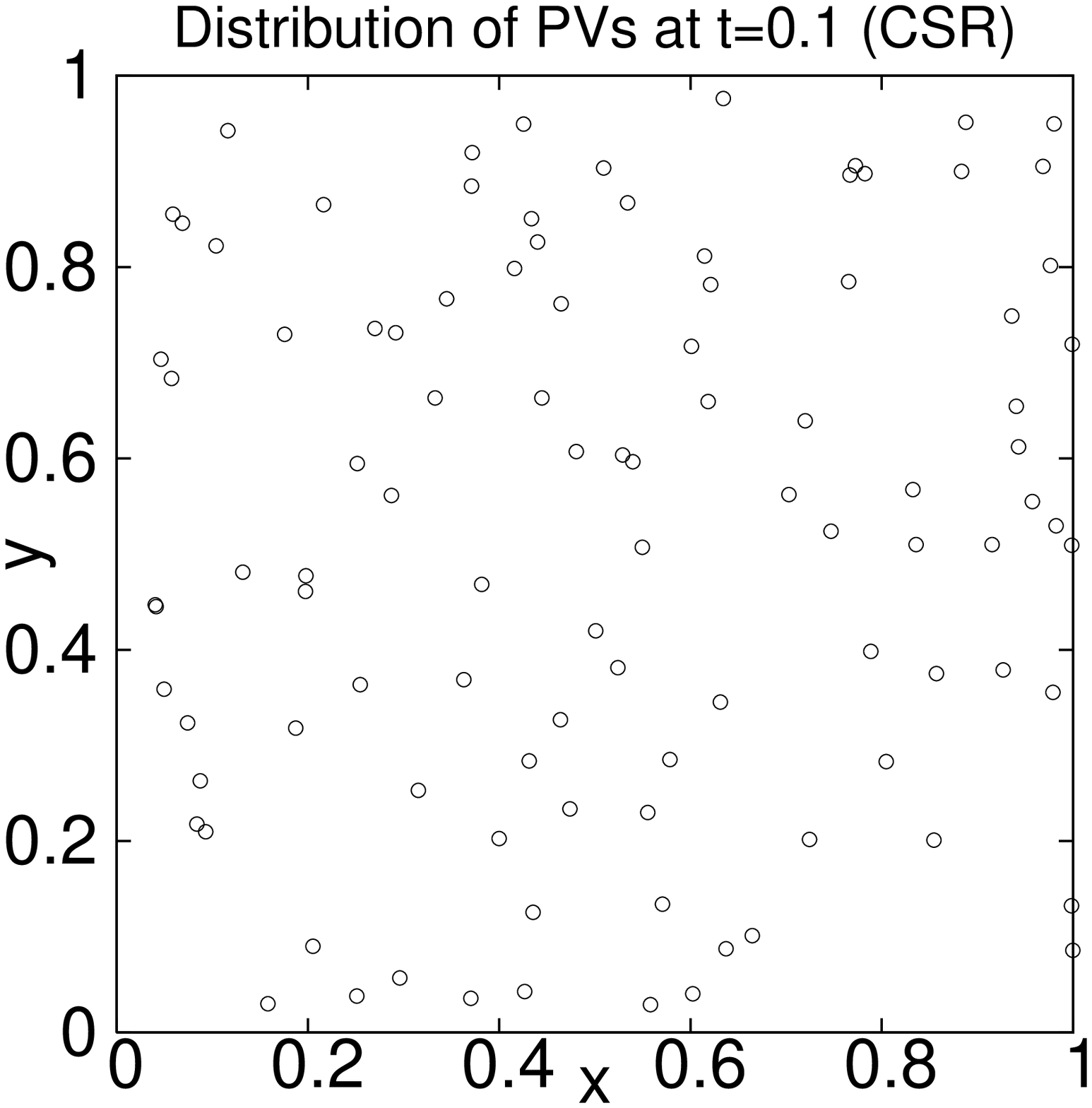}} \\
    \end{tabular}
    \caption{(IV) The positions of PVs at $t=0$ and $t=0.1$, 
    denoting the CSR pattern.}
    \label{fig4}
  \end{center}
\end{figure}

In Case (I), we have an analytical expression of $L(r)$ 
for a single row. The roll-up of PVs starts immediately 
corresponding to the Kelvin-Helmholtz instability. 
$L>0$ for the row and it persists at the final time $t=t_f$. 
The relative errors of the conserved quantities are confirmed to be 
below $10^{-5}$ at the final time in all numerical examples. 

\section{Numerical simulation of two types of PVs}

Next, we consider the three cases of positive and negative PVs, 
(V) the K\'arm\'an vortex street, 
(VI) positive and negative PVs located alternately 
in checkered segments (16 subsquares), and
(VII) the CSR distribution. 
We can define similarly the $K_{lm}$ and $L_{lm}$ functions 
for $(l,m) = (+,+), (-,-),$ and $(+,-)$, where 
$+ (-) $ denotes the positive (negative) PVs. 

In Case(V), the pairing of positive and negative PVs appears. 
This corresponds to the pairing instability of the double row 
of the staggered PVs \cite{Saffman}. 
The typical final time can be estimated as 10 $t_e$, where $t_e$ 
is the average eddy turnover time \cite{Umeki}. 

Clustering of both types of PVs is remarkable again in Case (VII). 
The vortex pair moves linearly at a velocity proportional to $1/h$ 
where $h$ is the distance of two PVs. 
Then, the pair is scattered by a third vortex, or recoupling occurs. 
This type of scattering on the unbounded plane was solved analytically 
by Aref \cite{Aref}. 
We made a GIF animation of motions of PVs. 
The motion is quite interesting. 
It looks like a motion of gas molecules, which does not 
exist in the case of the single type of PVs. 

Another method to quantify the distribution of PVs in squares of the same size 
\cite{Umeki2}
is called the Quadrat method \cite{Cressie} in spatial ecology. 

\begin{figure}[H]
  \begin{center}
  \vspace{-25mm}
    \begin{tabular}{cc}
      \resizebox{60mm}{!}{\includegraphics{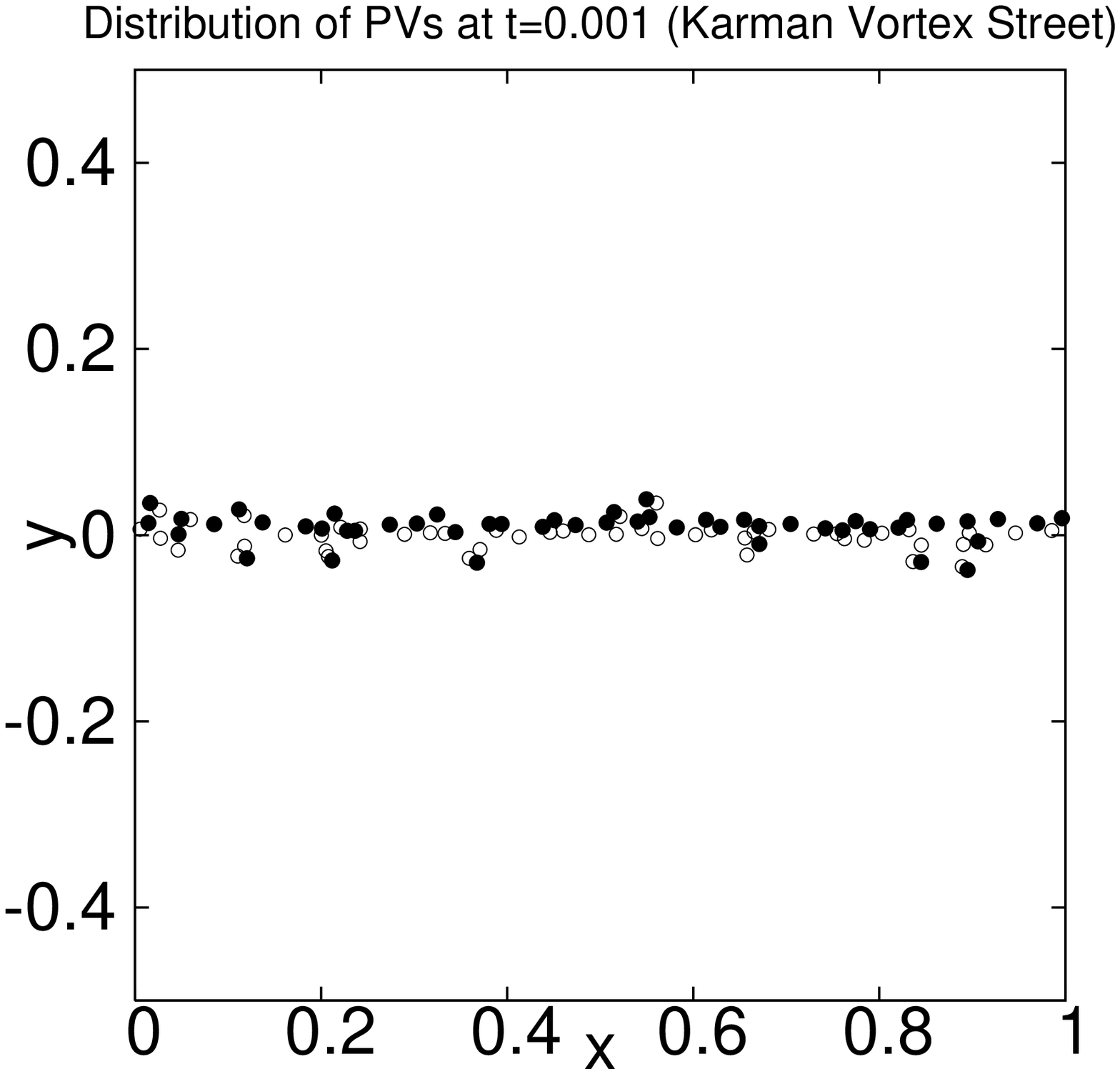}} &
      \resizebox{60mm}{!}{\includegraphics{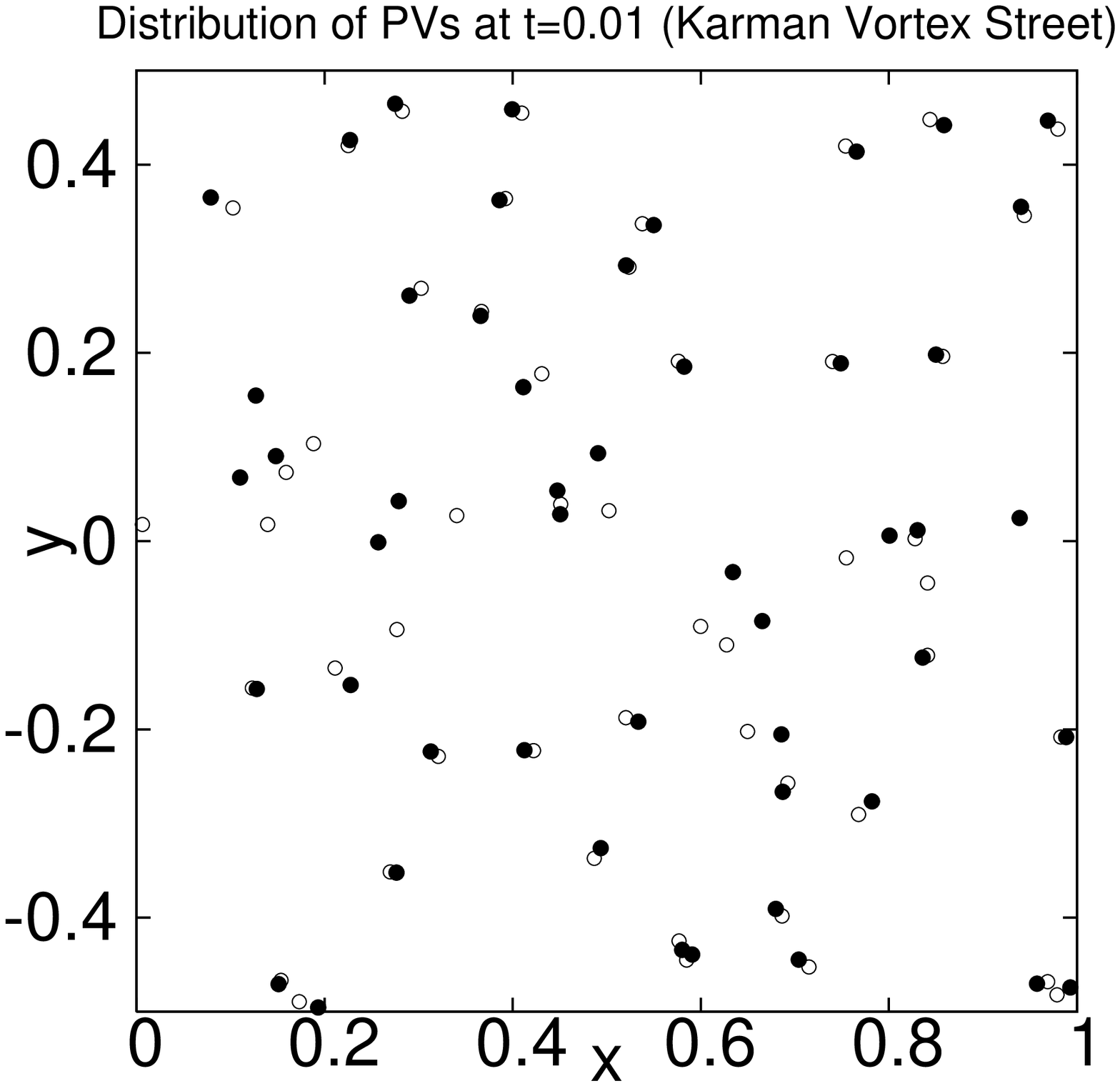}} \\
    \end{tabular}
    \caption{
(V) The positions of positive and negative PVs at $t=0.001$ and $t=0.01$, 
denoting the model of K\'arm\'an vortex street. Black (white) circles denote positive 
(negative) PVs. }
    \label{fig8}
  \end{center}
  \vspace{-25mm}
  \begin{center}
    \begin{tabular}{cc}
      \resizebox{60mm}{!}{\includegraphics{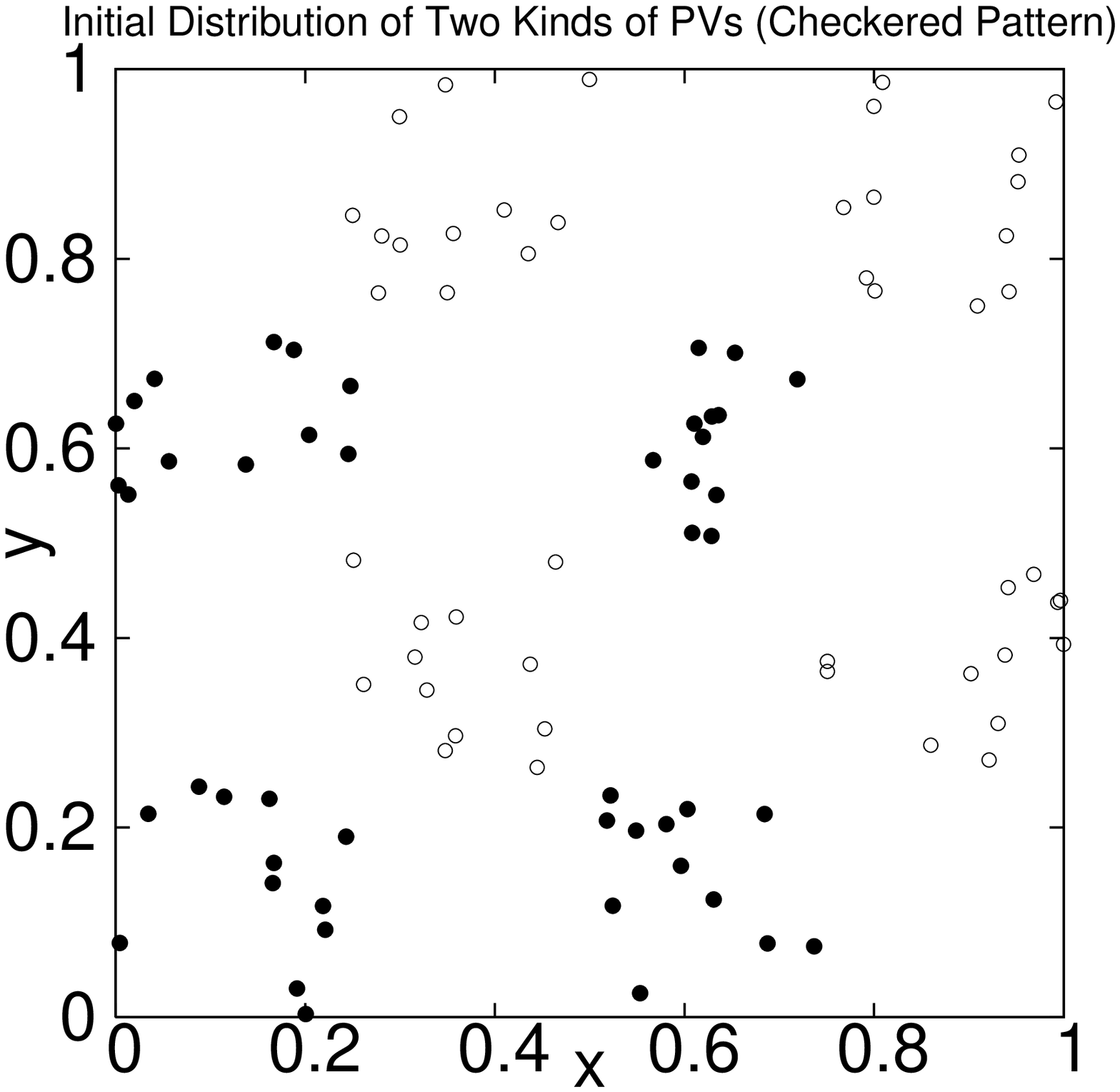}} &
      \resizebox{60mm}{!}{\includegraphics{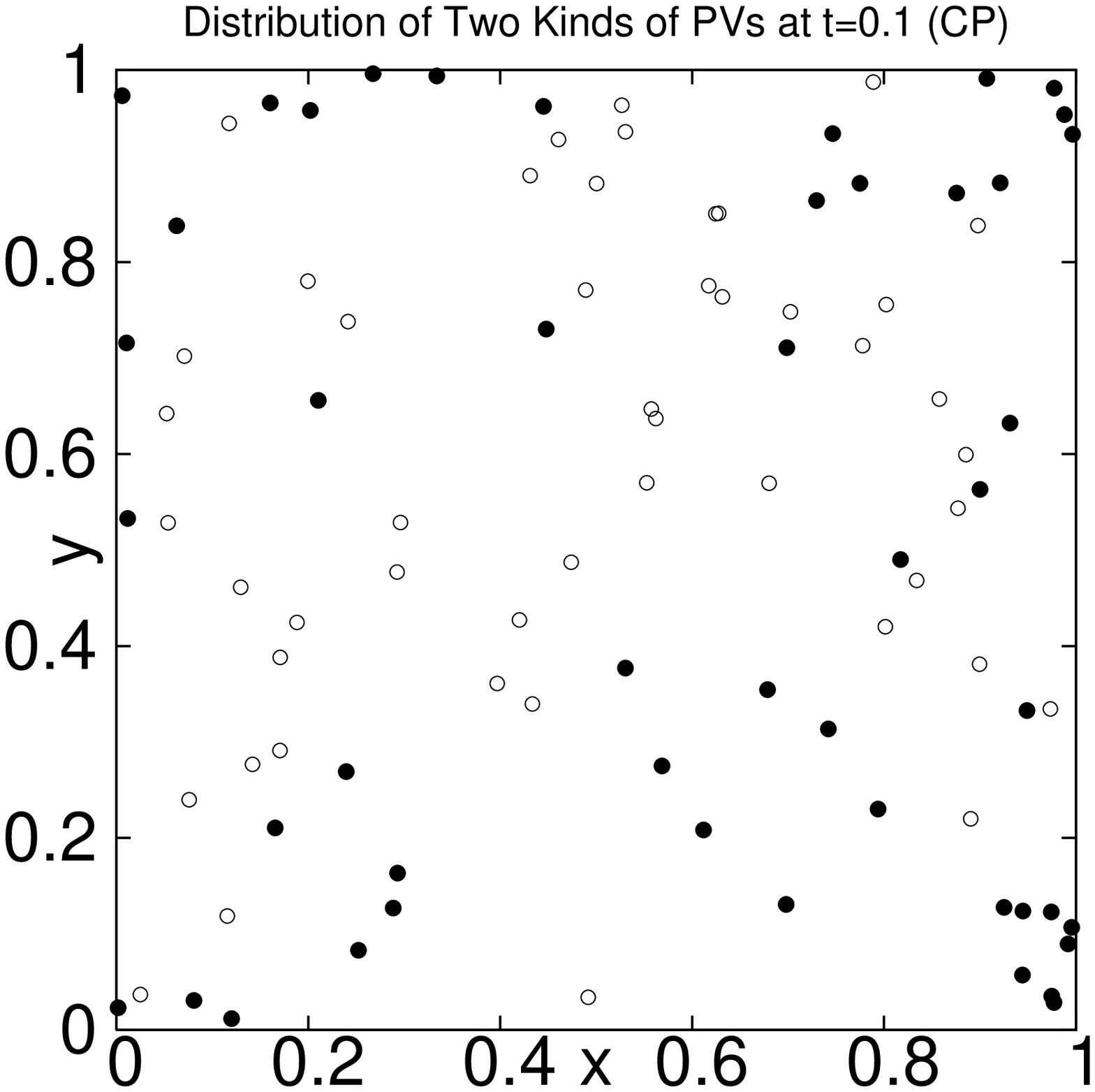}} \\
    \end{tabular}
    \caption{(VI) The positions of positive and negative PVs at $t=0.001$ and $t=0.1$, 
    denoting the checkered pattern. }
    \label{fig9}
    \end{center}
\vspace{-25mm}
  \begin{center}
    \begin{tabular}{cc}
      \resizebox{60mm}{!}{\includegraphics{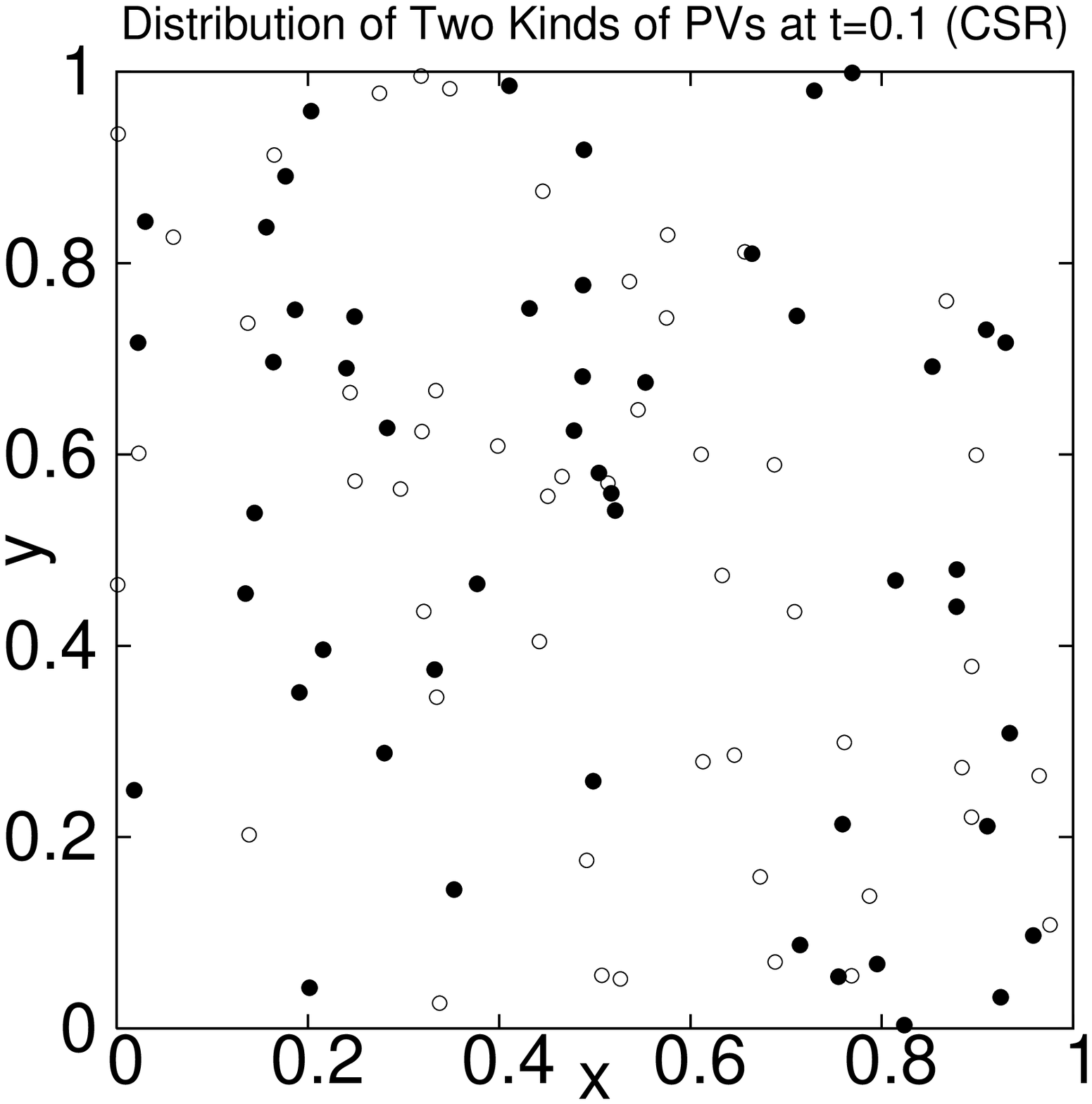}} &
      \resizebox{60mm}{!}{\includegraphics{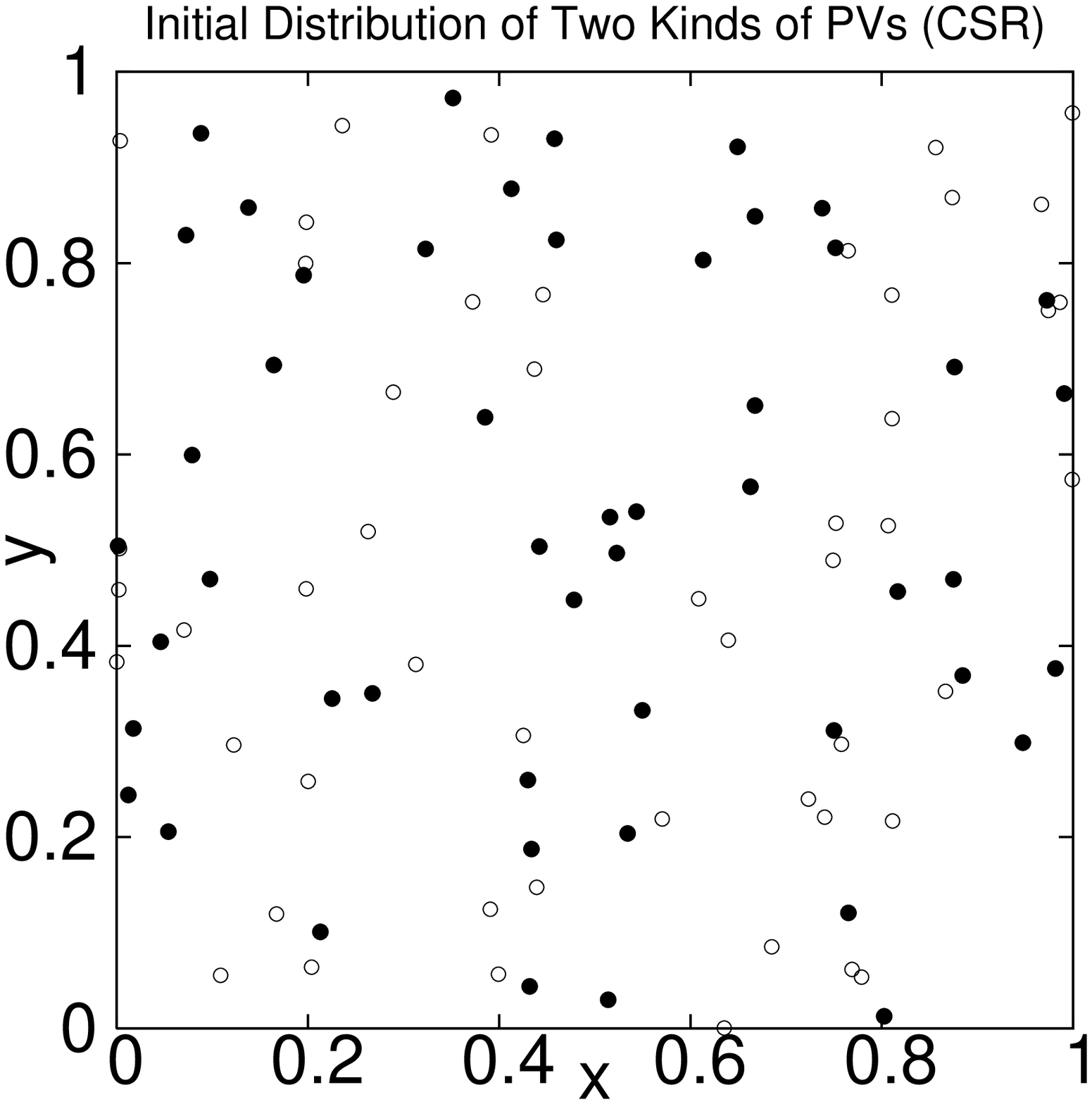}} \\
    \end{tabular}
    \caption{(VII) The positions of positive and negative PVs at $t=0.001$ and $t=0.1$, 
    denoting the CSR distribution.}
    \label{fig10}
      \end{center}
\end{figure}

\section{Conclusions and Discussions}
Clustering of PVs is studied numerically based on point process theory and
the $L$ functions are computed from the final distributions of PVs. 
Clustering persists up to 10 eddy turnover time. 
We also did a numerical simulation for slightly perturbed 
square vortex lattices, which can be regarded as a special circumstance 
of Case (III). It showed inactive motions of PVs compared with the 
other examples shown in this paper. It may imply that the instability 
of square lattices requires more than 100 PVs, or the rate of growth 
is very small. The author is grateful for Professor T. Yamagata for support through his 
research on fluid dynamics over these several years.

\end{document}